\documentclass[structabstract]{aa}  
\usepackage[authoryear]{natbib}
\usepackage{graphicx}
\usepackage{txfonts}

\def\smallsun{\hbox{$_\odot$}}

\def\arcsec{\hbox{$^{\prime\prime}$}}
\def\cm3{cm$^{-3}$}

\begin{document}

\title{Planetary nebulae abundances and stellar evolution II\thanks{Based
   on observations with ISO, an ESA project with instruments funded
   by ESA Member States (especially the PI countries: France,
   Germany, the Netherlands and the United Kingdom) and with the
   participation of ISAS and NASA.}}

\author{S.R.\,Pottasch\inst{1} \and J.\,Bernard-Salas\inst{2}  }

\offprints{pottasch@astro.rug.nl}

\institute{Kapteyn Astronomical Institute, P.O. Box 800, NL 9700 AV
 Groningen, the Netherlands \and Center for Radiophysics and Space
 Research, Cornell University, Ithaca, NY-14850-6801, USA}

\date{Received date /Accepted date}

\abstract
{In recent years mid- and far infrared spectra of planetary 
nebulae have been analysed and lead to more accurate abundances. It may be 
expected that these better abundances lead to a better understanding of the 
evolution of these objects.}
{The observed abundances in planetary nebulae are compared to those predicted 
by the models of Karakas (2003) in order to predict the progenitor masses of
the various PNe used. The morphology of the PNe is included in the comparison.
Since the central stars play an important role in the evolution, it is expected
that this comparison will yield additional information about them.}          
{First the nitrogen/oxygen ratio is discussed with
relation to the helium/hydrogen ratio. The progenitor mass for each PNe can be
found by a comparison with the models of Karakas. Then the present luminosity 
of the central stars is determined in two ways: first by computing the central
star effective temperature and radius, and second by computing the nebular
luminosity from the hydrogen and helium lines. This luminosity is also a 
function of the initial mass so that these two values of initial mass can be
compared.}
{Six of the seven bipolar nebulae can be identified as descendants of high 
mass stars (4M\smallsun - 6M\smallsun) while the seventh is ambiguous. Most of 
the elliptical PNe have 
central stars which descend from low initial mass stars, although there are
a few caveats which are discussed. There is no observational evidence for a
higher mass for central stars which have a high carbon/oxygen ratio.
The evidence provided by the abundance comparison with the models of Karakas
is consistent with the HR diagram to which it is compared. In the course of 
this  discussion it is shown how `optically thin' nebulae can be separated 
from those which are 'optically thick'.}
{}




\maketitle

\section{Introduction}

Planetary nebulae (hereafter PNe) are an advanced stage of stellar
evolution of low and intermediate mass stars. After the asymptotic giant 
branch (AGB) phase is completed, these stars evolve through the PN stage before
ending their lives as white dwarfs. The gaseous nebula seen now as PN is the 
remnant of the deep convective envelope which once surrounded the core. This
core is now seen as the central star of the PN. The present abundances in the 
nebula reveal information about the chemical processes that took place during 
the AGB. These processes, which have first been discussed by Iben \& Renzini 
\cite{iben} and Renzini \& Voli \cite{renzini}, change the abundances according
to the mass of the star involved and the initial abundances in the star. Thus 
by investigating the PN abundances it may be possible to assign an initial mass
to the star.

Models have been made of the evolution of stars of different masses. These were
initiated with the discussion of Paczynski \cite{pacz} followed by the 
detailed calculations of Sch\"{o}nberner \cite{schon}, Vassiliadis and Wood
\cite{wood} and Bl\"{o}cker \cite{blocker}. These models refer mostly to post
AGB evolution. Models referring to evolution on the AGB have been made by 
several authors, e.g. Marigo et al. \cite{marigo} and Karakas \cite{karakas}. 
The latter models predict changes in the chemical composition which have 
occurred during the evolution and which have been brought to the surface and
subsequently expelled as the nebula. It is these models which will be used to
compare with observed PN abundances because not only do they follow a star of
a given mass over its entire life, but the same is done for an entire sequence 
of possible masses for several different initial abundances.

The purpose of the present paper is to compare these models with the
abundances which we have observed. These abundances have been
determined with the help of mid and far infrared observations either from {\em
 ISO} or {\em Spitzer} and are quite accurate because they are less
affected by possible temperature variations or gradients in the nebula. These
observations have already been used (Pottasch \& Bernard-Salas\,\cite{pobs} to
better determine PN abundance gradients in the galaxy.  In an ideal
case it might be expected that a comparison of models with
observations will lead to: 1) knowledge of the individual properties
of the central stars, and 2) confirmation or suggestion for
improvement of the models.  In practice these goals are rather
difficult to reach because of shortcomings of both the observations as
well as the models. On the observational side are uncertainties in the
effective temperature of the central star, their distances, as well as
the accuracy of the measurements. The models presently available are
uncertain because the physical conditions in the actual star-nebula
system is poorly known. For example, the mass loss along the AGB (and
post AGB) is physically not well understood and the initial conditions
may not be realistic. Thus models used for comparison are taken from
different authors who may use different mass loss rates. Therefore core masses
are used where possible although initial masses are given for the Karakas 
models because the author identifies them as such.

The abundances observed are listed in Table 1. No indication is given there of 
the spectrum of the central star. A few of these stars are Wolf-Rayet stars for
which it may be that some of the evolutionary calculations may not apply. These
are the central stars of BD+30 3639, NGC\,40, NGC\,5315 and NGC\,6369. This can
be kept in mind when making the comparisons. 

The objects were selected to be IR bright (in the diaphragm of the instrument 
used). This was first done with the ISO spectrometer where almost all of the 
usable PN spectra were investigated. Later the Spitzer IR spectra of PNe have 
been investigated. Most of these spectra are as bright or nearly as bright as
the ISO PNe. This may at first suggest a bias toward PNe with massive central 
stars because these initially evolve at the highest luminosity. But the period 
of high luminosity is expected to be very short so that very few, if any, high
mass central star PNe are expected. We therefore may expect that many low mass
central star PNe have been observed, not only because of much longer evolution 
time but also because of the much greater number of low mass objects present. 
It is expected that most of the observed PNe are reasonably local objects, 
within a few kpc of the sun.

Nevertheless a confrontation of the models with the observations, even with 
these limitations, can give interesting insights into the evolution of the PN
system. In Sect.\,2 the morphology of the nebulae will be discussed, first in 
relation to the nitrogen/oxygen ratio observed in the nebula, and then the
helium abundance will be introduced into the discussion. In Sect.\,3 the 
effective temperature of the central star and the various ways of determining
it will be discussed. Then the luminosity of the central stars will be 
discussed. Because the luminosity is dependent on the distance of the nebula
this will also be discussed in this section. In addition a digression will be 
made into the long-standing question of whether or not a nebula is 'thick' to
photons which ionize hydrogen. This can be done because two different methods
of obtaining the nebular luminosity are available, one which makes no 
assumption concerning the nebular 'thickness' and one which is dependent on
this assumption. In Sect.\,4 we investigate whether the expected relation 
between abundance and luminosity can be seen. Conclusions and discussion are
given in the last section.

\section{Morphology and abundance}

There is a long history of discussion of the nebular morphology and its
relation with the nebular abundance. In 1971 Greig \cite{greig1} classified
the large majority of PN in two categories depending on their shape. He 
called these categories B (binebulous) and C (centric). He noted that the B
nebulae have stronger forbidden lines of \ion{N}{ii} and \ion{O}{ii} although 
he did not directly relate this to abundance. Greig \cite{greig2} also noted
that the B nebulae have kinematic properties indicating that they are the
younger group. Several years later Peimbert \cite{peimbert} classified the
PN solely on the basis of the nebular abundance in four classes. His type I 
nebulae are nitrogen and helium rich while type II are an intermediate
population having nitrogen and helium abundances close to solar. Type III, 
called `high velocity' does not have a substantially
different composition as type II while type IV is a poorly understood group
called `halo' PN which is small in number and will not concern us here.

In the five years that followed it became clear that those nebulae classified 
as type I also showed morphological similarities. This is summarized in an
article by Peimbert \& Torres-Peimbert \cite{peimbert2} where many, but
certainly not all, the PN listed as type I show morphology given as filamentary
and bipolar. The use of abundance to indicate morphology, for example by 
calling a nebula type I on the basis of an abundance determination, a practice 
which for some years was common, led to false morphological classification. The
morphological classifications listed in column 10 of Table 1 are based on 
detailed study, generally based on optical photographs, 
which is summarized by Manchado \cite{manchado}. We also adopt the system
which he gives dividing PN into three classes: bipolar (B), elliptical (E) and
round (R). This system is not universally used, probably because it is not
obvious that there is a fundamental difference between elliptical and round.
The classification given is taken mostly from Phillips \cite{phillips} who
takes this mostly, but not always, from earlier discussions in the literature. 
Because not all our nebulae are listed by Phillips we have also used other
sources (e.g. Stanghellini et al. \cite{stang} \& Manchado et al. \cite{man2}).
There is general agreement as to the classification for more than 90\% of the 
PN but there are cases of disagreement. Stanghellini et al. have classified
IC\,4191, NGC\,5315 and NGC\,6369 irregular, but we follow all other observers
(e.g. Phillips \cite{phillips} in calling them E or R. There is also 
speculation in the literature concerning the uncertainty introduced by 
projection effects but we have not attempted to include this uncertainty.

\subsection{Nitrogen/Oxygen ratio}

As discussed above, we have investigated the abundances in a large number of
PNe using the mid and far infrared observations from {\em ISO} or {\em Spitzer}.The resulting abundances are summarized in Table 1, which has been taken mostly
from Pottasch \& Bernard-Salas \cite{pobs} but include a few new results. We
regard these results as more accurate than other abundances found in the 
literature and have the advantage that they have been derived in the same way, 
which is why we have only used this sample.

A histogram of the nitrogen/oxygen (N/O) ratio for the 33 PN listed in Table 1 
is plotted as Fig.1. Six of the seven bipolar nebulae have a very high log N/O 
ratio, between 0 and 0.4.  The single exception to this is NGC\,6445 which 
clearly has a lower value. Of the five round nebulae (not plotted), four have 
values of log 
N/O less than -0.66. The fifth nebula (NGC\,5315) has a much higher log N/O 
ratio (-0.05), closer to that of the bipolar PN. All the elliptical nebulae 
have log N/O less than -0.8, but mostly in the range -1.0 to -0.4, with the 
exception of M1-42. This is a rather weak nebula and has been less well studied
than the other PN. Thus we confirm the correlation between shape and N/O ratio
and show that the morphology seems to change at a log N/O ratio of about -0.1.

\begin{figure}
 \centering
 \includegraphics[width=9cm]{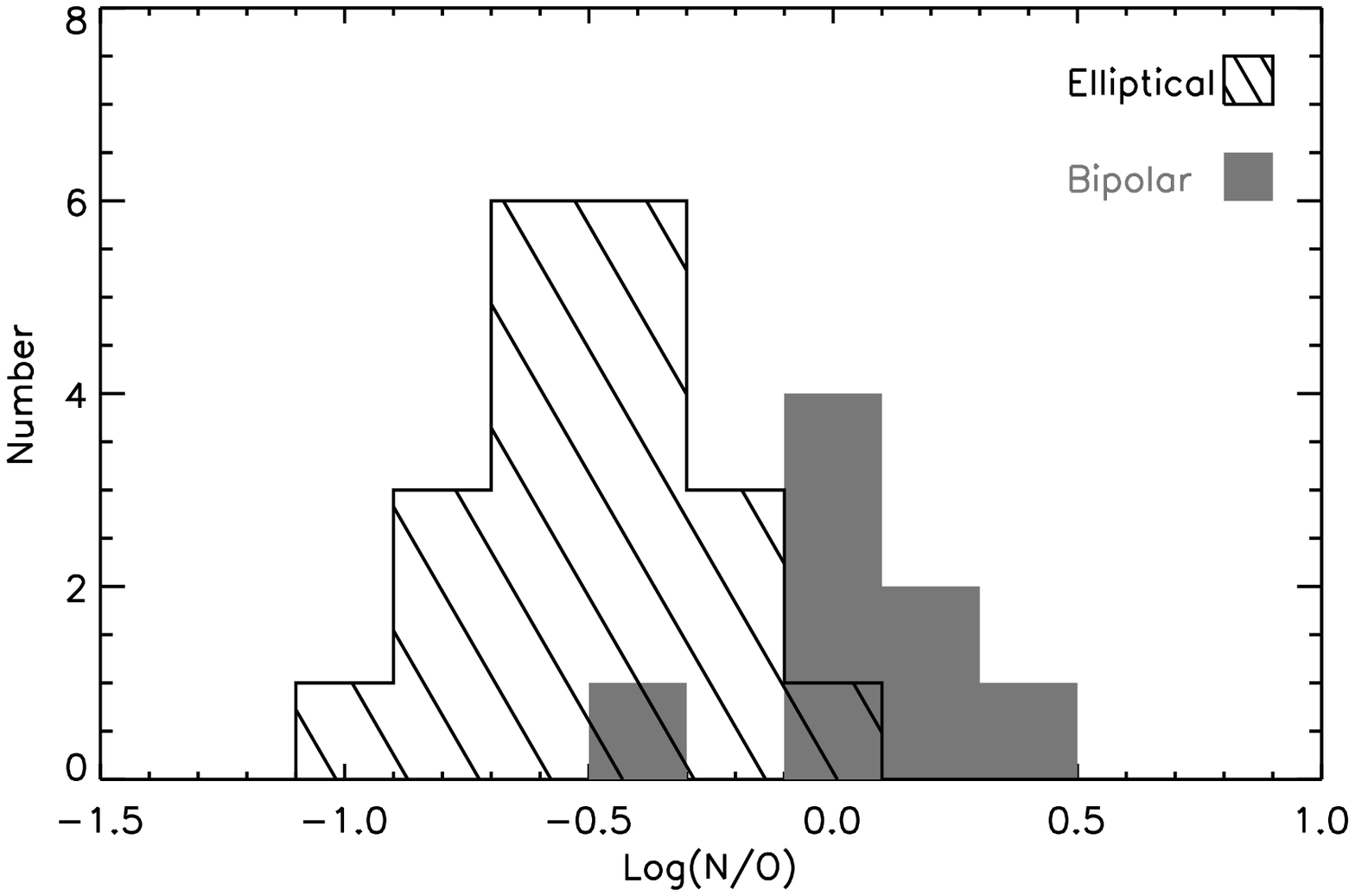}
 \caption{Histogram of the nitrogen/oxygen ratio of the PNe listed in Table 1.
Round PNe are not included in this plot because of the small number of objects.}
 \label{fig-1}
\end{figure}

\begin{table*}[!ht]
\caption[]{Elemental abundance of PNe with far-infrared data in addition to 
optical and UV data.}
\begin{center}
\begin{tabular}{|l| c c c c c c c|  c c c  |}
\hline
\hline
PNe & He/H & C/H & N/H & O/H & Ne/H & S/H & Ar/H & R$^\natural$ & Morph &  Ref.\\
& &$\times$10$^{-4}$&$\times$10$^{-4}$&$\times$10$^{-4}$&$\times$10$^{-4}$&$\times$10$^{-5}$&$\times$10$^{-6}$& (kpc) &   &\\
\hline

BD+30\,3639 &         &  7.3  & 1.1  &  4.6  &  1.9  &  0.64 &  5.2    & 7.6  & R & a   \\      
Hb\,5       &   0.123 &       & 4.5  &  4.5  &  2.2  &  0.70 &  5.5    & 6.0  & B & b   \\
He\,2-111   &   0.185 &  1.1  & 3.0  &  2.7  &  1.6  &  1.5  &  5.5    & 6.2  & B & c   \\
Hu\,1-2     &   0.127 &  1.6  & 1.9  &  1.6  &  0.49 &  0.42 &  1.1    & 7.9  & B & d   \\   
IC\,418     &  $>$0.072 &  6.2& 0.95 &  3.5  &  0.88 &  0.44 &  1.8    & 8.8  & E & e   \\
IC\,2165    &   0.104 &  4.8  & 0.73 &  2.5  &  0.57 &  0.45 &  1.2    & 9.8  & E & e, $\ast$ \\
IC\,4191    &   0.123 &       & 1.49 &  7.7  &  4.7  &  1.6  &  4.45   & 7.0  & E & p   \\ 
M\,1-42     &   0.161 & 10.5  & 7.5  &  8.3  &  4.4  &  2.8  &  8.6    & 3.0  & E & h   \\ 
Me\,2-1     &   0.1   &  7.0  & 0.51 &  5.3  &  0.93 &  0.91 &  1.6    & 5.8  & R & r   \\
Mz\,3       &  $>$0.080 & $<$16.  & 3.0  &  2.3  &  1.2  &  1.0  &  5.0 & 6.3 & B & q \\ 
NGC\,40     &  $>$0.046 &  19 & 1.3  &  5.3  &  1.4  &  0.56 &  3.4    & 7.9  & E & f   \\
NGC\,2022   &   0.106 &  3.66 & 0.99 &  4.74 &  1.34 &  0.63 &  2.7    & 9.4  & E & p   \\
NGC\,2440   &   0.119 &  7.2  & 4.4  &  3.8  &  1.1  &  0.47 &  3.2    & 8.9  & B & g   \\     
NGC\,5315   &   0.124 &  4.4  & 4.6  &  5.2  &  1.6  &  1.2  &  4.6    & 6.3  & R & i   \\
NGC\,5882   &   0.108 &  2.2  & 1.6  &  4.8  &  1.5  &  1.3  &  2.9    & 7.2  & E & e, $\ast$ \\
NGC\,6153   &   0.140 &  6.8  & 4.8  &  8.3  &  3.1  &  1.9  &  8.5    & 6.9  & E & f   \\
NGC\,6302   &   0.170 &  0.6  & 2.9  &  2.3  &  2.2  &  0.78 &  6.0    & 6.4  & B & j   \\
NGC\,6445   &   0.151 &  7.4: & 2.4  &  7.4  &  2.0  &  0.79 &  3.8    & 5.8  & B & k   \\
NGC\,6537   &   0.149 &  1.7  & 4.5  &  1.8  &  1.7  &  1.1  &  4.1    & 6.0  & B & c   \\ 
NGC\,6543   &   0.118 &  2.5  & 2.3  &  5.5  &  1.9  &  1.3  &  4.2    & 8.1  & E & a   \\
NGC\,6741   &   0.110 &  6.4  & 2.8  &  6.6  &  1.8  &  1.1  &  4.9    & 6.5  & E & l   \\      
NGC\,6818   &   0.099 &  5.4  & 1.26 &  4.8  &  1.27 &  0.94 &  2.7    & 6.6  & E & p   \\
NGC\,6886   &   0.107 &  14.3 & 4.2  &  6.5  &  2.0  &  1.0  &  2.1    & 7.7  & E & o   \\
NGC\,7027   &   0.106 &  5.2  & 1.5  &  4.1  &  1.0  &  0.94 &  2.3    & 7.4  & E & m   \\
NGC\,7662   &   0.088 &  3.6  & 0.67 &  4.2  &  0.64 &  0.66 &  2.1    & 8.2  & E & l   \\
IC\,2448    &   0.094 &  2.7  & 0.55 &  2.5  &  0.64 &  0.20 &  1.2    & 8.0  & E & s   \\ 
NGC\,2392   &   0.080 &  3.3  & 1.85 &  2.9  &  0.85 &  0.50 &  2.2    & 8.4  & E & t   \\
NGC\,6826   &   0.10  &  4.8  & 0.58 &  3.95 &  1.5  &  0.26 &  1.4    & 8.0  & E & u   \\
NGC\,3242   &   0.092 &  1.95 & 1.35 &  3.8  &  0.90 &  0.28 &  1.7    & 8.1  & E & v   \\
NGC\,6369   &   0.102 &   -   & 0.79 &  4.0  &  0.79 &  0.60 &  1.6    & 7.2  & R & v   \\
NGC\,6210   &   0.092 &  1.2  & 0.70 &  4.9  &  1.2  &  0.74 &  2.3    & 7.0  & R & w   \\
NGC\,2792   &   0.103 &  1.1  & 0.68 &  2.1  &  0.74 &  0.43 &  1.6    & 8.2  & E & x   \\
NGC\,1535   &   0.085 &  1.6  & 0.21 &  2.6  &  0.53 &  0.13 &  1.1    & 9.9  & E & z   \\

\hline

\end{tabular}
\end{center}
$^\natural${Galactocentric distance assuming the Sun is at 8\,kpc from the center},
Morphology: R=round, E=elliptical, B=bipolar\\ 
$^\ast${Higher resolution observations.}\\ 
{{\em References:} a) Bernard-Salas et al. 2003, A\&A
406, 165, b) Pottasch et al. 2006, c) Pottasch et al. 2000, A\&A, 363,
767, d) Pottasch et al. 2003, A\&A, 401, 205, e) Pottasch et al. 2004,
A\&A, 423, 593, f) Pottasch et al. 2003, A\&A, 409, 599, g)
Bernard-Salas et al. 2002, A\&A, 387, 301, h) Pottasch et al. 2007, A\&A, 471,
865, i) Pottasch et al. 2002, A\&A, 393, 285, j) Pottasch et
al. 1999, A\&A, 347, 975, k) van Hoof et al. 2000, ApJ, 532, 384, l)
Pottasch et al. 2001, A\&A, 380, 684, m) Bernard-Salas et al. 2001,
A\&A 367, 949, o) Pottasch \& Surendiranath 2005, A\&A, 432, 139, p)
Pottasch et al. 2005, 436, 965, q) Pottasch \& Surendiranath 2005,
A\&A, 444, 861, r) Surendiranath et al. 2004, A\&A, 421, 1051, s) Guiles, S.
et al. 2007, ApJ, 660, 1282, t) Pottasch et al. 2008, A\&A, 481, 393,
u) Surendiranath \& Pottasch 2008, A\&A, 483, 519, v) Pottasch \& 
Bernard-Salas 2008, A\&A 490, 715, w) Pottasch et al. 2009, A\&A, 499, 249,
x) Pottasch et al. 2009, A\&A, 502, 189, z) unpublished. } 

\end{table*}

\subsection{Helium vs. N/O}

The helium, nitrogen and oxygen abundances with respect to hydrogen are listed 
in Table 1. The values of N/O are plotted in Fig.2 as a function of He/H. The
elliptical PNe are plotted as diamonds and the bipolar PNe as asterisks. It 
can be seen that six of the seven bipolar nebulae lie on the upper right
hand side of the figure, while NGC\,6445 lies somewhat by itself with a rather
low N/O ratio but a rather high value of He/H. The two elliptical PNe which
have a high N/O ratio and thus lie close to the six bipolar nebulae are M1-42
and NGC\,5315. The abundances predicted by Karakas \cite{karakas} for stars of 
different initial masses are labeled with numbers in the figure indicating the 
masses used. These masses are connected with lines for each of the three 
different values of heavy element abundance Z.

\begin{figure}
\centering
 \includegraphics[width=9.0cm]{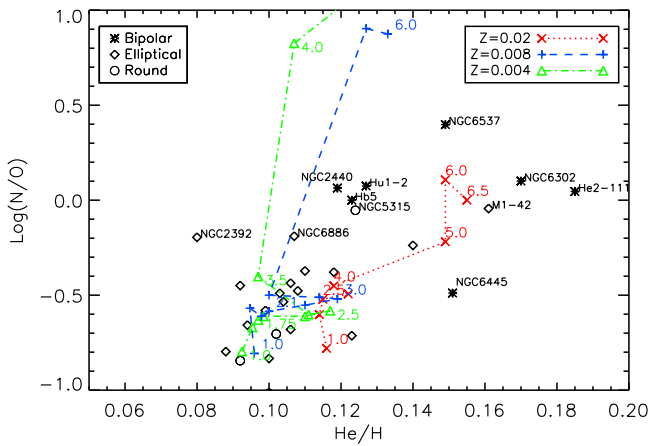}
 \caption{The N/O abundance ratio is plotted as a function of He/H. The 
dashed lines connect the 
results of the models of Karakas for a given value of Z. The ciphers give the 
initial masses of the individual models. The individual measured PN abundances
are from Table 1. Those referred to in the text also have their names given.}
 \label{fig-2}
\end{figure}

An individual comparison between the observed abundances and those predicted by
Karakas is difficult because the initial values of helium and the heavy 
elements (Z) is not known for the PNe. The value of Z can be partially computed
from the observed abundances but several important elements are not observed.
Even for observed elements such as oxygen and carbon it is possible (although
unlikely) that important 
amounts are tied up in dust and therefore not observable. But still important
conclusions can be drawn from Fig.\,2. First of all, the observed points appear
to lie between the curves for Z=0.02 and 0.008, i.e. there is a general 
agreement between the predicted and observed abundances. Four PNe seem to be 
in the range of PNe with initial mass greater
than 5\,M\smallsun. Three of them are bipolar nebulae: NGC\,6302, NGC\,6537 and
He2-111. They appear to have passed the stage of hot-bottom burning.  The 
fourth PN, M1-42, is difficult to compare with the theoretical curves because
the high abundances indicate that it may have a high value of Z, and perhaps
a high initial helium abundance since it lies much closer to the
galactic center than the other PNe. Three other bipolar PNe, NGC\,2440, Hb5 and
Hu1-2 appear to have a slightly lower initial mass, between 4M\,\smallsun~and
5\,M\smallsun, a value of Z=0.008 or somewhat higher, and have also have 
undergone hot-bottom burning, which destroys carbon to produce nitrogen and 
possibly some oxygen as well. The elliptical nebula
NGC\,5315 is also in this category. The three elliptical PNe NGC\,6153, 
NGC\,6886 and NGC\,2392 have rather high N/O ratios indicating initial masses
slightly above 4\,M\smallsun. They have different He/H ratios however which
could indicate that they are stars of different initial helium abundances. All
the other PNe are very close to the Z=0.008 curve for stars between initial 
masses of 1\,M\smallsun~and 4\,M\smallsun, and are all elliptical PNe. The only
exception is the bipolar PN NGC\,6445, whose N/O ratio indicate that it is in
this initial mass range but it has a much higher He/H ratio and therefore 
difficult to understand. The large majority of the nebulae can be
interpreted with central star masses which agree with the helium and nitrogen
abundances predicted by Karakas.

\subsection{The Carbon abundance}

The question now arises whether the observed carbon abundances fit
into this picture. Carbon abundances are somewhat more uncertain than
nitrogen abundances because all the observable ions are in the
ultraviolet which makes them much more dependent on correct extinction and
electron temperature determination. We lack four determinations
of carbon in these nebulae because of the very large extinction
in the ultraviolet spectrum of these nebulae which made it impossible to
measure the carbon lines.  In
Fig.\,3 the N/O ratio has been plotted against the C/O ratio for those PNe
with carbon abundances. The predicted values of these ratios
(Karakas \cite{karakas}) as a function of stellar mass for initial
values of Z$=$0.004, 0.008 and 0.02 are shown as points connected by dashed
lines.  It may be seen that a very similar picture emerges as that drawn from 
the N/O vs He/H plot. The only three
PNe which are close to the 5\,M\smallsun~and 6\,M\smallsun~lines
are again the bipolar nebulae NGC\,6302, NGC\,6537 and He2-111. M1-42
now is separated from this group and is in the neighborhood of the
two other bipolar PNe Hu1-2 and NGC\,2440. In Hb5 the extinction is
too high for the carbon lines to be measured. The last bipolar nebula,
NGC\,6445 is again at a position of lower initial mass. Again
NGC\,5315 and NGC\,6153 are at a position predicted for a mass of
4.5\,M\smallsun, as are NGC\,2392 and NGC\,6886. The low helium abundance of 
these two PNe placed them in a
more anomalous position in Fig.\,2, but both the N/O and C/O ratios
indicate that they are evolved from stars of initial mass somewhat
more than 4\,M\smallsun.  NGC\,6741 has about this mass as well.

\begin{figure}
 \centering
 \includegraphics[width=9.0cm,height=9cm]{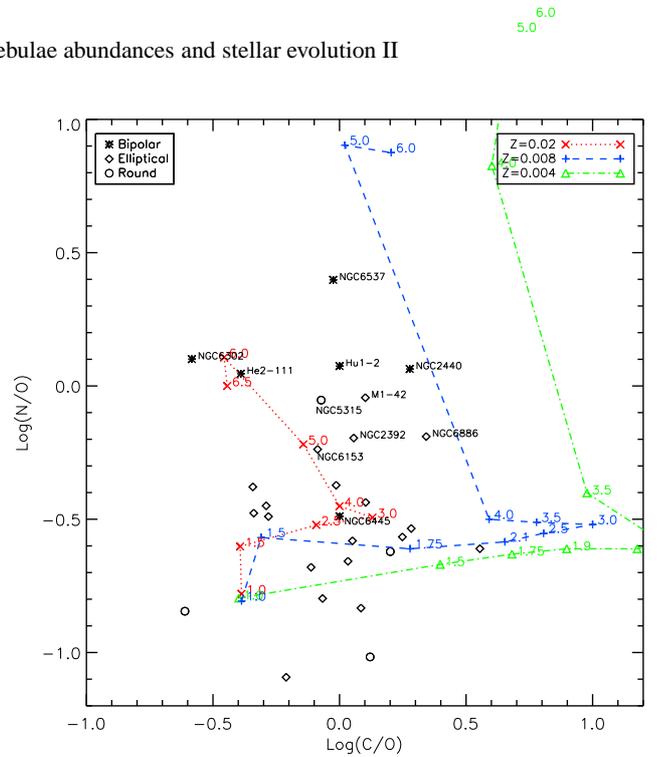}
 \caption{N/O plotted against C/O. The lines and points are the same as in the
caption to Fig.2.}
 \label{fig-3}
\end{figure}

The remaining PNe have lower initial mass. When the C/O ratio greater
than unity and the log N/O ratio is less than -0.46, the initial mass is
probably between 2\,M\smallsun~and 3.5\,M\smallsun. Probably the
nebulae IC\,418, IC\,2165, NGC\,40, NGC\,7027, IC\,2448, Me\,2-1,
BD+30\,3639 and NGC\,6826 are in this category.  A log C/O ratio lower
than -0.15 combined with a log  N/O ratio less than -0.46 indicates an
initial mass less than 2\,M\smallsun. The extreme example is the
central star of NGC\,6210 which probably has the smallest initial mass (equal 
to or less than unity).  Because the initial stellar abundances are not known
this should only be taken as an indication.

\section{Stellar Mass and Luminosity}

After the thermally pulsing AGB phase terminates, low to intermediate mass 
stars evolve at near constant luminosity to higher temperature. The ejected 
envelope becomes ionized and it is seen as a planetary nebula. Models made of 
single star evolution through this phase (Sch\"{o}nberner \cite{schon}, 
Vassiliadis \& Wood \cite{wood}, Bl\"{o}cker \cite{blocker}) show that the 
constant stellar luminosity is a strong function of the core mass of the
star. Thus determination of the luminosity provides a second method of
determining the initial stellar mass which may be compared to the mass found 
from the abundances in the previous section. This is the purpose of the 
present section. First the effective temperature will be discussed, then the 
radius of the central star will be found using the visual magnitude of the
central star. The distance of the PN plays an important role in determining
the luminosity and a subsection is devoted to this subject.

The stellar luminosity can also be determined by measurements of the planetary 
nebula alone, even when the central star is not visible. The measurement used 
is either the amount of radiation which will ionize hydrogen or the amount of
radiation which will doubly ionize helium. For the first case all the hydrogen
ionizing radiation must be measured i.e. no ionizing radiation can escape from
the nebula without causing at least one hydrogen ionization. For the second 
case this must be true for radiation which doubly ionizes helium. The 
luminosity determined in this way can be directly compared with that determined
from the measurement of the central star. This is because both methods have the
same dependence on the PN distance so that the distance is unimportant in the 
comparison. Interesting conclusions can be drawn about the 'optical thickness'
of the nebula as well.

\subsection{Stellar temperature}

The effective temperature of the central stars can be determined from the 
Zanstra method providing the spectral distribution of radiation resembles that
of a blackbody and that all the photons which are able to ionize both hydrogen 
and ionized helium are actually absorbed, i.e. the nebula is `optically deep'
for these radiation fields. This is because the Zanstra method compares the 
total amount of ionizing radiation (every photon which ionizes hydrogen 
produces a single H$\beta$ photon and the H$\beta$ flux is measured) with the
flux in the visible part of the spectrum (as measured by the stellar visual
magnitude). Consider first the stellar magnitude determination.

\subsubsection{Stellar magnitude}

The difficulty in measuring the magnitude of the central star is the result of 
the presence of 
nebular emission in the diaphragm. This must be large enough so that the star 
remains in the diaphragm. This is especially difficult when the star is very
faint in the visible. In recent years this situation has improved: Hubble Space
Telescope (HST) measurements have become available so that the image of the 
star is no longer broadened by the seeing. For most of the faint central stars 
magnitudes from the HST are now available. These are listed in the second 
column of Table 4. Theseare measured values and they must still be corrected 
for extinction. The  extinction constant C is listed in col.\,3 and the 
reference in col.\,4 of the table. The accuracy can be judged by comparing the 
two central stars for which multiple  measurements are available; these are 
shown in Table 2,

\begin{table}[htbp]
 \caption[]{Comparison of measured stellar magnitudes.}
 \begin{center}
 \begin{tabular}{lrcc}
 \hline
 \hline
 Nebula    &      (1)    &    (2)  &     (3)\\
 NGC\,2440  &    17.63   &   17.49  &   17.6\\
 NGC\,7027  &    16.04   &   16.53  &   16.18\\
 \hline
 \end{tabular}
 \end{center}
\end{table}

where col.(1) are the HST measurements of Wolff et al. \cite{wolff}, col.(2) 
are HST measurements of Ciardullo et al. \cite{ciar} and col.(3) are the ground
based measurements made by Heap \& Hintzen \cite{heapa}. Very faint central
stars are difficult to measure. We list the HST measurement of Matsuura et al.
\cite{mats} for NGC\,6537 in Table 4 but for NGC\,6302 and He2-111 the central 
stars are too faint or too obscured to permit measurement.

\subsubsection{Stellar radius}

Once the stellar magnitude is known, the stellar radius can be determined if 
the distance is known. This uncertain quantity will be discussed presently; The
distances used are listed in col.10 of Table 4. Further it is assumed that the
stars radiate as blackbodies so that the following equation can be used:

\begin{equation}
\left(\frac{R_{S}}{d}\right)^{2} = 4.808 \times 10^{-18} \times
(e^{\frac{1.439}{\lambda T}}-1) \times 10^{\frac{m_{vo}}{2.5}} 
\end{equation}

where d is the distance, $\lambda$ is 5480\AA, T is the stellar temperature
and m$_{vo}$ is the magnitude 
after correction for extinction. The resultant stellar radius is listed in 
col. 11 of Table 4. The stellar temperature used is listed in col.8 of Table
4 and is discussed below. It does not have a large effect on the value of
the radius.

\subsubsection{Zanstra temperature}

The Zanstra temperature measures the ratio of the amount of ionizing radiation
to the amount of radiation in the visual spectrum.  T$_{Z}$(H) uses the 
H$\beta$ flux and thus measures the amount of radiation which can ionize 
hydrogen, while  T$_{Z}$(HeII) measures the amount of radiation which can 
completely ionize helium. In converting the ratio to a temperature it is 
assumed that the stellar spectrum is a blackbody, that all the ionizing 
radiation is absorbed by the nebula, and that the continuum visual flux is 
measured by the stellar magnitude.  

Zanstra temperatures have been measured for many years and several
papers have published extensive tables of these temperatures (e.g.
Phillips \cite{phillips}).  The most uncertain measurement in the
determination is the stellar magnitude because, as discussed above,
nebular light must be avoided. The values we have found are listed in
cols. 5 and 6 of Table 4 and are not essentially different from those
given in the literature. One of the most discussed aspects of the
results can be seen in the table: in about 30\% of the nebulae the
value of T$_{Z}$(H) is substantially lower than T$_{Z}$(HeII). The
reason for this has been debated in the literature; the most often
cited reason is the assumption that the nebula is 'optically deep' to
radiation which ionizes hydrogen is  wrong and that some of this radiation
escapes the nebula without being registered. This has the consequence
that T$_{Z}$(H) is too low and that T$_{Z}$(HeII) is the more nearly
correct value. It is difficult to confirm this because not only is the
total nebular mass uncertain, its distribution in the nebula is also
unknown. Another explanation for this difference could also be that
the stellar spectrum is not well represented by a blackbody.

\subsubsection{Energy balance temperature}

The Energy Balance method, first introduced by Stoy \cite{stoy}, measures the 
average excess energy per ionizing photon. This can be found from the ratio of
the intensity of collisionally excited nebular lines to H$\beta$. It has the 
advantage that only the nebular spectrum has to be known; no measurement of the
central star flux is necessary. It has the further advantage that it is 
applicable both to optically thin as well as optically thick nebulae. The 
method is also independent of the nebular model as long as all the 
collisionally excited lines are measured. In practice sometimes a correction 
must be made for unmeasured lines. The entire spectrum must be measured but for
most PN the visible and ultraviolet lines are the most important. For very low
temperature central stars the infrared nebular spectrum can be the most 
important. 

Once the ratio of collisional line intensity to H$\beta$ (called R) is known a 
difficulty arises in interpreting this measurement in terms of a stellar 
effective temperature; it is necessary to know whether the star emits as a 
blackbody or some particular model atmosphere. Since this is not known it is 
assumed here that the star emits as a blackbody. Preite-Martinez \& Pottasch 
\cite{pmp} have calculated the effective temperatures found when a variety of 
model atmospheres of different effective temperatures are used as ionizing 
source. They found that for a fixed value of the ratio R the model atmospheres 
give a slightly lower value of effective temperature than the blackbody.

The exact status of the nebula also has an effect on the effective temperature
found. Preite-Martinez \& Pottasch \cite{pmp} calculated three cases. In the 
first case the nebula is optically thin to all ionizing radiation. In the
second case the nebula is optically thick to He$^+$ ionizing radiation and in 
the third case the nebula is optically thick to all ionizing radiation. These
authors compare the effective temperatures derived for 52 central star using
all of these three assumptions. They find that all three assumptions give
similar results. We have redone the calculations using the case which is thick
to He$^+$ ionizing radiation and thin to hydrogen ionizing radiation (case 
two); the effective temperatures found are listed in col.8 of Table 4. As can
be seen from the table, temperatures can now be found even when the central
star is unobservable. The Energy balance temperature is rather similar to
the HeII Zanstra temperature T$_{Z}$(HeII), sometimes slightly lower, 
sometimes slightly higher.

\subsubsection{Spectroscopic temperatures}

Stellar temperatures may also be obtained from a model atmosphere analysis of
the spectrum. Because high resolution spectra are needed this has only been 
done for very bright stars. The results can be found in Mendez et al. \cite{mendez}, Kudritzki et al. \cite{kudritzki} and Pauldrach et al. \cite{pauldrach}. 
As Pauldrach et al. \cite{pauldrach} point out, the model atmosphere analysis 
is difficult; the 
results using hydrogen line profiles can differ according to which hydrogen 
line is used. Mendez et al. \cite{mendez} \& Kudritzki et al. \cite{kudritzki}
base their temperatures on the analysis of hydrogen and helium line profiles
while Pauldrach et al. \cite{pauldrach} base their temperatures on the analysis
of metal line profiles in the ultraviolet. The results are quite similar. The
results are given in Table 3 where, considering the consistency of the
different determinations, we estimate the error to be of the order of 10 to 
15\%.

Although only six spectroscopic temperatures have been measured for our 
nebulae, it is interesting to compare them with what has been found from the 
Zanstra and Energy Balance methods. For two of the nebulae, IC\,418 and 
NGC\,6826, no  T$_{Z}$(HeII) can be measured. In both cases there is good 
agreement between the spectroscopic temperature and  T$_{eff}$ derived from
T$_{Z}$(H) and  T$_{EB}$. In two other cases, NGC\,3242 and NGC\,1535, there
is reasonably good agreement between the spectroscopic temperature and  
T$_{eff}$ derived from T$_{Z}$(HeII) and  T$_{EB}$, but definitely higher than
that found from T$_{Z}$(H). This could also be true for IC\,2448 because the 
spectroscopic temperature is more uncertain for this central star. It is 
definitely not true for the central star of NGC\,2392 where both  T$_{Z}$(HeII)
and  T$_{EB}$ indicate a very much higher temperature. This will presently be
discussed in more detail.

\begin{table}[h]
\caption[]{Spectroscopic temperatures.}
\begin{center}
\begin{tabular}{lrcc}
\hline
\hline
Nebula &   T$_{eff}$(1) &    T$_{eff}$(2)&    T$_{eff}$(3) \\
\hline
NGC\,2392  &  40,000    &       45,000   &      47,000 \\
NGC\,3242  &  75,000    &       75,000    &     75,000 \\
IC\,418   &   39,000   &        37,000   &      36,000 \\
NGC\,6826  &  44,000   &        50,000   &            \\
NGC\,1535  &           &                 &      70,000 \\
IC\,2448    &          &                 &      65,000 \\

\hline

\end{tabular}
\end{center}
(1)Pauldrach et al. \cite{pauldrach}; (2)Kudritzki et al. \cite{kudritzki};
(3)Mendez et al. \cite{mendez}

\end{table}

\subsubsection{Effective temperature}

In col.8 of Table 4 the estimated value of the effective temperature  T$_{eff}$
is given. For those PNe where both  T$_{Z}$(HeII) and  T$_{EB}$ have been 
measured, they are always quite similar. Of these 27 cases the average value of
T$_{eff}$ is similar (within 15\%) to the hydrogen Zanstra temperature
T$_{Z}$(H) in 15 cases. In the other 12 cases T$_{Z}$(H) is considerably
lower than T$_{eff}$, possibly indicating that the nebulae are optically thin.
We shall return to this subject in the next section. Four of the nebulae are 
not hot enough to form a substantial amount of ionized helium. In these cases
substantial weight is given to T$_{EB}$ which in two cases is very similar to
T$_{Z}$(H). In the other two cases T$_{EB}$ is higher than T$_{Z}$(H) and an
average of the two temperatures is used. For Mz\,3 the temperature is 
determined from the best fitting nebular model (see Pottasch \& Surendiranath
\cite{ps1}). The value of  T$_{eff}$ for NGC\,6302 is discussed below.

\begin{table*}[!ht]
\caption[]{Central star Magnitude, Temperature and Luminosity.}
\begin{center}
\begin{tabular}{|l| c c c| c c c c| c  c  c | c c c|}
\hline
\hline
PNe & m$_{V}$ & C & Ref. & T$_{Z}$(H) &T$_{Z}$(HeII) &T$_{EB}$ &T$_{eff}$ &H$\beta$ & dist & R$_{S}$/R\smallsun &    &L$_{S}$/L\smallsun& \\
   & meas. &  (3)  & (4)  &  (5)    & (6)    & (7)    & (8)    & x10$^{11}$ & kpc & (11)   &   (12)  &  (13) &   (14)   \\
\hline

BD+30       &  10.28  &  0.50  &  9  &  31,500  &         & 30,000  & 30,000    & 29.5  & 1.0 & 1.42  & 1470 & 2450 &   \\      
Hb\,5       &  17.1   &  1.60  &  8  & 105,000  & 130,000  & 145,000 & 140,000  & 11.9  & 2.3 & 0.128  & 5600 & 2700 & 4550  \\
He\,2-111   &         &        &     &         &           &          & 200,000: & 1.34 & 2.5 &     &      &      & 650   \\
Hu\,1-2     &  17.76  &  0.60  &  6  & 100,000 & 145,000  & 130,000  & 140,000 &  2.22  & 1.5 & 0.0335  & 390  & 210 & 540  \\   
IC\,418     &  10.23  &  0.33  &  1  &  34,000  &       &  36,000   &   35,000  & 58.3  & 1.0 &1.3  & 1850  & 3700 &     \\
IC\,2165    &  17.47  &  0.59  &  2  & 132,00   & 135,000 & 120,000 &  134,000  & 4.98  & 1.5 & 0.0774  & 1750 & 1810 & 2680 \\
IC\,4191    &  16.4   &  0.48  & 11  &  79,000  &  90,000  & 97,000  &  90,000  & 5.25  & 2.1 & 0.099  & 580 & 750 & 1300   \\ 
M\,1-42     &  17.4   &  0.63  &  h  & 63,000   &  81,000  & 100,000 &  85,000  &  1.0  & 5.0 & 0.179  & 1500 & 810 & 1690  \\ 
Me\,2-1     &  18.40   & 0.28  &  2  & 130,000  & 145,000 & 142,000  & 145,000  & 0.935 & 2.3 & 0.0273 & 300  & 210 & 480  \\
Mz\,3       &         &  1.65.  &    &         &         &         &  39,500    & 29.6  & 1.3 &        &      & 2300 &     \\ 
NGC\,40     &  11.55  &  0.605  & 1  &  35,000  &        & 40,000 &  38,000    & 17.3   & 0.8 & 0.683  & 875  & 570  &    \\
NGC\,2022   &  15.75 &  0.38   & 1,11  &  59,000 &  108,000 & 105,000 & 108,000  & 1.79   & 1.8 & 0.099 & 1200  & 220  & 1500   \\
NGC\,2440   &  17.63 &  0.50  & 2    &  208,000  &  205,000  & 180,00 & 200,000  & 9.1   & 2.0  & 0.037 & 2000 & 1900 & 2100   \\     
NGC\,5315   &  14.3  &  0.54  & 3    &  65,000  &  76,000  &  66,000  & 70,000  & 13.3 & 2.0  & 0.306  & 2020 & 1720 &    \\
NGC\,5882   &  13.42 &  0.33  & 1    &  50,000  &  64,000  &  70,000  & 67,000  & 12.7 & 1.2  & 0.231 & 979  & 590 & 1070 \\
NGC\,6153   &  15.55 &  1.14  & 1    &  77,000  &  87,000  &  80,000  &  82,000  & 21.5 & 1.2  & 0.169 & 1200 & 1000 & 2000 \\
NGC\,6302   &        &        &      &        &           & 300,000:  & 300.000:  & 56.  &  1.6  &       &       & 8000 & 7400  \\
NGC\,6445   &  18.72 &  1.06:  & 3   &  185,000  & 175,000  &  175,000 & 180,000  & 8.7  & 1.8 & 0.0348 & 1200 & 1380  & 1170   \\
NGC\,6537   &  21.6  &  1.79  & 5   &  410,000  &  350,000  &        & 350,000:  & 12.8 & 2.0  & 0.0147 & 2950  & 4200 & 3700   \\ 
NGC\,6543   &  11.29 &  0.10  &  1  &  47,000  &         & 58,000    & 56,000    & 30.8  & 1.0  & 0.457 & 1840  & 1050 &    \\
NGC\,6741   &  20.09 &  1.1  &   4  & 205,000  & 219,000  & 230,000  & 220,000  &  4.1  & 1.9 & 0.184   & 710 & 850 & 400   \\      
NGC\,6818   &  17.02 &  0.35  &  3 &  140,000  & 155,000 & 138,000 &  145,000  & 7.43  & 2.0 & 0.048 & 940 & 1270 & 2150   \\
NGC\,6886   &  18.76 &  0.70  &  6  &  145,000  & 145,000  & 156,000  & 152,000 & 2.34  & 2.6  & 0.038  & 680 & 670  & 690   \\
NGC\,7027   &  16.07 &  1.3  & 2,10  & 165,000  & 160,000  & 234,000 & 180,000  & 134.  & 0.9  & 0.0716  & 4820  & 5300 & 4100   \\
NGC\,7662   &  14.00 &  0.18  & 1   &  80,000  & 108,000   & 95,000  & 103,000  & 15.3  & 1.2  & 0.118 & 1400  & 770  & 2500  \\
IC\,2448    &  14.26 &  0.27  &  1   &  48,000  & 83,000    & 91,400  &  85,000  & 2.35  & 1.4  & 0.147 & 1160  &  151 & 850  \\ 
NGC\,2392   &  10.63 &  0.22  & 1   &  37,000  & 78,000   & 80,000  & 80,000   & 6.83  & 1.5    & 0.845 & 26000 & 490 & 4450  \\
NGC\,6826   &  10.68  & 0.07  & 1   &  34,000  &        & 45,000    & 42,000   & 12.8  & 1.4    & 0.98  & 2680  & 1120 &    \\
NGC\,3242   &  12.32  & 0.12  & 12   &  57,000  & 90,000  & 70,000  & 80,000  & 21.4  &  0.55   & 0.14  &  740  & 260 & 1600   \\
NGC\,6369   &  15.91  & 2.12   & 3,12 & 69,000  & 71,000  & 70,000  & 70,000  & 61.6  &  1.2   & 0.40  & 3250  & 2850  & 2900  \\
NGC\,6210   &  12.66  & 0.14  & 7   &  51,000  & 61,000   & 69,000  & 65,000  & 11.0  & 1.57   & 0.36  & 2070  & 890  & 1060   \\
NGC\,2792   &  16.89  & 0.80  & 1,11  & 82,000  & 135,000  & 126,000  & 130,000  & 3.05  & 1.9 & 0.079 & 1600 & 430  & 1570   \\
NGC\,1535   &  12.11  & 0.08  &  1   &  40,000  &  76,000  & 87,000  & 80,000    & 4.8  &  2.1   & 0.519   & 9900   & 680   & 2500    \\

\hline

\end{tabular}
\end{center}

{\em References:} 1)Ciardullo et al. \cite{ciar}, 2)Wolff et al. \cite{wolff} 
3)Gathier \& Pottasch \cite{gathier} 4)Sabbadin et al. \cite{sabbadin} 
5)Matsuura et al.  \cite{mats} 6)Heap et al. \cite{heapc} 7)Pottasch et al.
\cite{pbsr} 8)Tylenda et al. \cite{tylenda} 9)Crowther et al. \cite{crowther}
10)Zijlstra et al. \cite{zijlstra} 11)Preite-Martinez et al. \cite{pmak} 
12)Pottasch \& Bernard-Salas \cite{pobs}\\
H$\beta$ in units erg cm$^{-2}$ s$^{-1}$, Col.7 is the energy balance 
temperature, col.8 is the effective temperature, col.12 1s the luminosity found
from the magnitude and effective temperature of the central star, col.13 is the
luminosity found from the nebular hydrogen line, col.14 is the luminosity found
fron the nebulae ionized helium line.

\end{table*}

\subsection{PN Distances}

Distances to PNe are uncertain because the usual method for astronomical
distance determination, parallax, is applicable to only a very limited number
of nebulae. In this situation two options are available. One can use less
accurate methods for determining individual distances or one can assume that 
all nebulae have a particular property in common and use this property to 
obtain a statistical distance. There is an extensive literature for determining
distances with the assumption that all nebulae have the same ionized mass.
Especially the work of Cahn et al. \cite{cahn} or more recently the work of
Stanghellini et al. \cite{stang2}, which calibrates the mass using Magellanic 
Cloud PNe measurements, have been used. Both these determinations assume that
the PNe are optically thin to hydrogen ionizing radiation but attempt to
correct this assumption for the smaller high density nebulae.

The statistical distances will not be used here because PNe evolve from stars 
of a wide range of stellar masses and we do not wish to exclude the possibility
that PNe of different stellar mass produce nebulae with different properties.
Averaging any of these properties may lead to systematic errors in the 
individual distances found from these average properties. This in turn could 
introduce systematic errors in evolution calculations made using them. In 
addition Ciardullo et al. \cite{ciar} measured accurate distances of a small
sample of binary PNe and concluded that for these nebulae the statistical 
distances are overestimates. Instead we determine distances using three methods
which, while giving somewhat more uncertain individual distances, are much less
likely to have systematic errors.

These methods are well known and are long in the literature. The first method 
is the expansion distances where the expansion is measured at two epochs, 
usually separated by 3 to 5 years. The measurements are made in both the 
optical (HST measurements are desirable) and at radio frequencies (VLA
measurements are usually used). Expansion velocities must also be known but 
they cannot be measured in the plane of the sky in which the expansion is 
observed. The measured radial velocities are therefore used in the hope that 
these are very similar. In addition there may be gradients in the
velocity which must be taken into account. Because the nebular density 
decreases between the two measured epochs, the ionization front appears to move
more quickly than the matter and this must also be taken into account. A 
discussion with many results is given by Terzian \cite{terzian} and the effect 
of the ionization front is discussed by Mellema \cite{mellema}. In addition,
Sch\"{o}nberner et al.\cite{schon2} have made and applied kinematic models of
several PNe to obtain expansion distances. There is general agreement for the
four PNe common to the results of both Mellema\,\cite{mellema} and Sch\"{o}nberner et al.\cite{schon2}, which indicate some of the expansion distances given by
Terzian\,\cite{terzian} should be increased. The increases depend on the 
individual models and are thus uncertain. Both authors estimate that the 
results Terzian\,\cite{terzian} should on average increase by about 30\%. These
possible increases are taken into account in our distance estimates, although 
other methods, especially extinction distances, may place important limits on
the increase. For the PNe we are discussing 9 expansion distances are known 
(NGC3242, NGC6210, NGC6302, NGC6543, NGC7027, NGC7662, IC418, IC2448  and 
BD+30 3639).

The second method is the extinction distance. In this method the extinction of
field star of known distance located within a small distance from the nebula in
the plane of the sky (usually less than 0.5 degrees) is measured and a plot is 
made of reddening as a function of distance in that direction. The reddening of
the PNe is placed on this diagram and the distance read off. This method 
assumes that the extinction is rather uniform over the 0.5 degrees (or smaller)
used, and that the PNe is located in the galactic plane where the extinction
increases with distance.  This is a rather time consuming process. It
has been done for NGC2440, NGC2792, NGC5315, NGC5882, NGC6543, NGC6741, NGC7027
and IC2448 among the PNe we are interested in (see Gathier et al.
\cite{gath1}, Martin \cite{martin}).

There is a variant of this method which instead of measuring the extinction of 
the PN and nearby sources, measures the 21 cm neutral hydrogen radio line
absorption of the PN and nearby sources. This method has the advantage that the
velocity of the absorption line is measured so that it may be ascertained 
whether the nebula is on the near or far side of a particular spiral arm. This
method has been used for NGC6369, NGC6537, NGC6886 and NGC7027 (of the PNe 
which interest us). This method is again limited to PNe near the galactic 
plane (see Gathier et al. \cite{gath2}). 

In addition use can be made of the average extinction (in magnitudes per kpc)
in various directions. Such maps have been made by several authors but because
a relatively limited number of stars are used, these maps are averages over
relatively large areas and are more uncertain than when a small area is studied
in depth. Results of this method are given for a large number of PNe by
Pottasch \cite{pott12} and Sabbadin \cite{sabbadin2}. The results of these
two authors generally agree when the same nebulae are compared. We will use 
them when no other individual distance is available.

Another method of distance determination is from the spectroscopic measurements
of the central star. These measurements have already been discussed in section 
3.1.5 for determining the central star temperatures for which, with the 
exception of NGC2392, reasonable agreement was obtained with the temperatures
found from the nebular spectrum. The same discussions (Pauldrach et al.\cite{pauldrach}, Kudritzki et al.\cite{kudritzki} and Mendez et al.\cite{mendez}) 
derive the distance and mass of various bright central stars. These distances 
are not included in the present discussion. The reason for this is the 
following. The large masses and distances found by Pauldrach et al. are 
improbable. This has been convincingly demonstrated by 
Napiwotzki\,\cite{napiwotzki}, who found that the kinematic properties of these
nebulae are inconsistent with the masses and distances given by these authors. 
This reasoning can be extended to the distances found by Kudritzki et al. which
are very similar. Some, but not all, of the distances found by Mendez et al.\cite{mendez2} are similar to those given by Pauldrach et al. and Kudritzki et al.
There are six PNe in common between Mendez et al. and those used by us. For one
PN, NGC1535, the same distance is given. For two other nebulae, IC2448 and 
NGC3242 there is a strong discrepancy between the expansion distance as given 
by Mellema \cite{mellema} or Sch\"{o}nberner et al.\cite{schon2} in the sense 
that Mendez et al. give values of 3.5 and 1.8 kpc while the expansion distance 
is considerably lower, 2.1 and 0.55 kpc respectively. For the remaining 3 PNe,
NGC6826, NGC2392 and IC418 our individual distances are about 45\% smaller than
those given by Mendez et al.\cite{mendez2}. For the single PN for which an
optical parallax is available, NGC7293, Harris et al.\cite{harris} measure a 
distance of 219 pc while Mendez et al.\cite{mendez2} give the higher value of 
300 pc. We regard the values of Mendez et al. as rather high and do not give 
tham much weight.

Distances may also be determined from the nebular spectrum. When making a 
model of the nebula to explain, not only the relative intensity of the lines
relative to hydrogen, but the absolute hydrogen line intensities as well, the 
distance of the nebula is one of the unknown quantities (along with the stellar
radius and temperature and the nebular density and temperature). Our experience
with such models indicate that the distances determined in this way have an 
accuracy of about 20\%. Model distances are available for NGC2792, NGC6826,
NGC6886, NGC6741, NGC6445, Hu1-2, Hb5, Me2-1 and Mz3. In addition the distance 
to NGC1535 has been measured by the fact that it is a double star and the 
companion has a spectroscopic parallax.

In this way individual distances are known for 32 of the 33 PNe being studied.
The only exception is He2-111 for which a statistical distance has been used.
We feel that the distances listed in Table 4 are an improvement over the older
ones shown in our previous paper.

\subsubsection{Comparison with statistical distances}

In the above we have discussed the reason for not using the statistical 
distances. Here we compare the individual distances with the statistical 
distances found in the literature. There are two reasons for doing this. First,
to show that on average the statistical distances are not very different from 
the individual distances. This is to counter a possible remark when the nebular 
luminosities are discussed, that increasing PNe distances by a large amount is 
a reasonable alternative. Secondly we wish to demonstrate that bipolar PNe are
affected in a different way than elliptical PNe when statistical distances are
determined, thus demonstrating the bias discussed above.

There are many statistical distance scales in the literature and for clarity
only two of them will be discussed here. These are the often used scale of  
Cahn et al.\cite{cahn} (hereafter CKS) who calibrate distances using 'well 
known' PNe distances, and the more recent distances of Stanghellini et 
al.\cite{stang2} (hereafter SVV) who make use of the PNe in the Large 
Magellanic Cloud as calibrators. Both scales assume that all PNe are optically 
thin to hydrogen ionizing radiation and have the same ionized mass, but in 
both scales a correction to this assumption is made for nebulae considered 
optically thick. The differences between the distances given by these two 
sources is usually not large. 

Of the 32 PNe (excluding He2-111 for which no individual distance is
available), the individual distances to 20 nebulae are found to be the same as 
the statistical given by CKS, 8 PNe have larger individual distances and 4 PNe
have smaller individual distances. Compared to the distances given by SSV, 14
PNe have the same distance, 2 PNe are larger than given by SSV and 11 PNe have
smaller distances. Having the same distance is defined as being within 33\% of
the larger of the two distances. Note that in the comparison with SVV there are
slightly less PNe in common.

But if only the bipolar nebulae are considered a somewhat different picture 
emerges. Of the 7 bipolar nebulae (again excluding He2-111) the distances to 5 
of them are larger than given by CKS, 2 are at the same distance and there are 
no PNe at smaller distances. For SVV these numbers are 2, 3 and 0 respectively.
Thus while the individual distances are rather similar to the statistical 
distances for the whole sample, they are on the whole larger than the 
statistical distances for the bipolar nebulae. This
illustrates the danger of using statistical distances which averages out the
differences in different classes of nebulae. For the two PNe which on the basis
of the nebular abundance evolve from the most massive stars, NGC\,6537 and
NGC\,6302, we find distances twice as high as given by CKS or SSV.

\subsection{Luminosity} 

\subsubsection{Luminosity from central star.}

Once the effective temperature  T$_{eff}$, the stellar radius, and the distance
to the PNe are known the luminosity can be calculated. Using the relation:
\begin{equation}
L_S/L\smallsun=(R_S/R\smallsun)^2 \times (T_{eff}/T\smallsun)^4
\end{equation}
the stellar luminosity given in col.12 of Table 4 is found. It is difficult to
estimate the error of this luminosity. The distance is not well known so that 
the error could be as much as 30\%; for individual objects it might be even 
higher. This introduces an error of a factor of two in the luminosity and is
probably the largest error. The stellar magnitude and extinction value are
probably reasonably well known and are not likely to introduce an error of more
than 20\% unless the wrong star has been measured. We shall return to this
presently. The effective temperature, which is related to the assumption of
blackbody radiation, may also be in error. An error of 10\% in temperature
leads to an uncertainty of 40\% in luminosity.

\subsubsection{Luminosity from nebular emission}

The luminosity can also be determined from the emission lines of either 
hydrogen or ionized helium, at least when the nebula absorbs all the ionizing 
radiation. Consider hydrogen: every photon with energy greater than 13.6 eV
emitted by the central star ionizes one hydrogen atom and then produces a 
single H$\beta$ photon. Thus the H$\beta$ luminosity can be converted to the 
total luminosity of the star above 13.6 eV. When the temperature is known this
can be converted to the total stellar luminosity. This may be written as:

\begin{equation}
L_S/L_{H\beta}=C_1 \times T/G_1(T)                             
\end{equation}
where
\begin{equation}
C_{1}=\frac{\pi^4 k \alpha_{B(H)}}{15 h \nu_{H\beta} \alpha_{H\beta}}
\end{equation}

\begin{equation}
L_{H\beta}=4 \pi d^2 F_{H\beta}
\end{equation}

and G$_1$(T) is an integral shown as eq. VII-8 whose values are tabulated in
Table VII-4 by Pottasch\,\cite{pott12}. F$_{H\beta}$ is listed in column
9 of Table 4. The luminosity computed in this way has
already been used by Pottasch \& Acker\,\cite{pa}. It is similar to the Zanstra
method in the assumptions as well as in the equations. Like the Zanstra method
the luminosity can also be computed from the $\lambda$4686 \AA~line of ionized 
helium. The equations then become:

\begin{equation}
L_S/L_{4686}=C_2 T/G_4(T)                                  
\end{equation}

where

\begin{equation}
C_{2}=\frac{\pi^4 k \alpha_{B(He^+)}}{15 h \nu_{4686} \alpha_{4686}}
\end{equation}

\begin{equation}
L_{He4686}=4 \pi d^2 F_{4686} 
\end{equation}

where the references are the same as above.

The function  T/G$_x$(T) has a minimum value for both hydrogen and
ionized helium. For hydrogen the function does not change by more than 25\%
between temperatures of 45,000 K and 150,000 K. Within this range small errors
in the temperature will have only a reasonably small effect on the luminosity.
Outside of this range, and especially at the lower temperatures, an error in 
the temperature will have a much larger effect. This should be taken into 
account when comparing the luminosities determined in the different ways. For 
the luminosities found from the ionized helium line, the range of temperature
where the effect is small is between 140,000 K and 800,000 K. Again a small
error in temperature will have a much large effect at lower temperatures. This
is illustrated in the PN NGC6302. The central star, which is not visible, has
an uncertain  energy balance temperature of 300,000 K (Preite-Martinez \&
Pottasch\,\cite{pmp}). The luminosity computed from the $\lambda$4686 line is
7400L\smallsun. This value is only slightly dependent on the temperature 
between 200,000 K and 500,000 K. The value found from the H$\beta$ line is
much more sensitive to the temperature and only at a value close to 300,000 K
can the same luminosity be computed. This fixes the temperature.

The luminosities found from the H$\beta$ line are listed in column 13 of Table
4, and the luminosities found the $\lambda$4686 line are given in column 14 of
the same table. 

\subsubsection{Comparison of stellar and nebular luminosities}

Both the stellar and nebular luminosities have the same dependence on the 
distance so that a comparison of the two does not involve the distance. 

On making this comparison the general impression is of reasonably good 
agreement between
the stellar and nebular luminosities. On closer inspection we can distinguish
several cases. Case 1 are those PNe where the stellar luminosity (col.12 in
Table 4) and the $\lambda$4686 luminosity (col.14) agree and are both higher 
than theH$\beta$ luminosity. We regard these PNe as optically thin to hydrogen 
ionizingradiation and optically thick to radiation which doubly ionizes helium.
For these cases we use the average of col.12 and col.14 as the luminosity of 
the exciting star. There are eight cases: Hb5, Hu1-2, M1-42, Me2-1, NGC\,2022,
NGC\,5882, IC\,2448 and NGC\,2792. In case 2 are those nebulae where all three
luminosities agree. These are regarded as optically deep to both hydrogen and
doubly ionized helium. For these eight cases the average value of all three
luminosities is used. These are NGC\,2440, NGC\,5315, NGC\,6445, NGC\,6537,
NGC\,6886, NGC\,6369, NGC\,6741 and NGC\,7027. Then there are those nebulae 
(case 3) where the stellar luminosity agrees with that determined from the 
H$\beta$ luminosity, while that determined by the doubly ionized helium 
luminosity is about a factor of two higher. This is probably due to the use of 
a slightly too low effective temperature, since in all cases the temperature 
dependence of the luminosity is quite large. The PNe involved are IC\,2165, 
IC\,4191, NGC\,6153 and NGC\,6818. Here the stellar luminosity is used 
(col.12). These PNe are considered optically thick
to all ionizing radiation fields. It is likely that NGC\,7662 and NGC\,3242
should be included in this group because here again the doubly ionize helium
luminosity is about twice the stellar luminosity, but because these nebulae are
optically thin to hydrogen ionizing radiation the hydrogen luminosity is lower.
Case 4 are the five PNe which have low temperature central stars and thus do 
not have a HeII luminosity. Since these nebulae have low temperature central 
stars the nebular luminosity is a very strong function of the temperature. For
these five PNe (BD+30 3639, IC\,418, IC\,40, NGC\,6543 and NGC\,6826) we use
an average luminosity giving double weight to luminosity determined from the
star. 

We have now discussed 30 of the 33 PNe. For NGC\,6210 the stellar luminosity is
twice the nebular luminosity. We will use an average value of the three
luminosities listed. The problem 
concerning NGC\,2392 is greater. The stellar luminosity for this object is an 
order of magnitude greater than the nebular luminosity. It is also much greater
than for all the other nebulae. The problem probably lies with the temperature
we have assigned to this star. As discussed in Sec.\,3.1.5 and Table 4 the 
spectrum of this star is not more than 50,000 K, much less the value of 80,000
K which has been used. But a blackbody of 50,000 K does not have enough 
ultraviolet radiation to produce the observed nebular spectrum. Thus there must
be an additional source of ultraviolet radiation. This could be another hotter
star. This has been suggested several times in the literature because of the 
incompatibility of the nebular and stellar spectrum. Ciardullo et al.\cite{ciar} 
have examined HST photographs of this nebula and have found a faint star at
2.65\arcsec from the bright 'central star'. This star is only seen through the
red (I) filter; it is invisible in the visible (V). It is therefore likely to 
be a red star unless it has a very unusual spectrum. In any case, the stellar
luminosity listed for this star in Table 4 is certainly not correct. If one
wishes to assign a luminosity to the exciting star, we suggest using the
nebular luminosity found from the $\lambda$4686 line. The PNe is clearly
optically thin to radiation capable of ionizing hydrogen. 

The luminosities for NGC\,1535 show a great similarity to those of NGC\,2392.
Again the stellar luminosity is considerably higher than the nebular 
luminosities and the nebula is optically thin to radiation capable of ionizing 
hydrogen. But in this case the effective temperature which we have assigned to
the central star is not very different from the spectroscopic temperature shown
in Table 3 (although only a single determination is available). It is possible 
that in this case the nebula is optically thin to radiation capable of doubly 
ionizing helium. We find it difficult to assign a luminosity to this central
star.

There are 25 PNe which have both hydrogen and ionized helium Zanstra 
temperatures. We define the ratio T$_{Z}$(HeII)/T$_{Z}$(H) as the Zanstra
ratio. Ignoring NGC\,6210, there are 12 optically thick PNe and 12 optically
thin PNe as found above by a consideration of their luminosities. The Zanstra
ratio for the 12 optically thick PNe varies between 0.95 and 1.17 with a median
value of 1.03. For the 12 optically thin PNe the Zanstra ratio varies between
1.12 and 2.1 with a median value of 1.65. We conclude the assumption that the
star radiates as a blackbody is reasonable and consistent. Furthermore those 
PNe with a Zanstra ratio greater than 1.2 are optically thin, while those lower
than 1.1 are optically thick. Between values 1.1 and 1.2 further information is
necessary.

\subsection{Core mass and the HR diagram }

In the earlier discussions of the work of Karakas\,\cite{karakas} her various 
models have been identified by their initial mass, just as Karakas has done.
A more direct mass to use in discussing the HR diagram is the core mass, as it
avoids uncertainties in the assumed mass loss.  The Z=0.008 (Z=0.02) models of 
Karakas labeled with the initial masses of 1, 1.5, 2.5, 4.0 and 6.0M\smallsun~ 
have core mass values of 0.60(0.57), 0.63(0.60), 0.66(0.66), 0.84(0.79) and 
0.95(0.93)M\smallsun. The values of Z for the PN considered can be calculated
from the abundances in Table 1. The abundances are in general rather similar to
the solar abundances as given by Asplund et al.\cite{asplund}. These authors
calculate the solar valule of Z=0.0134. Thus most of the Z values calculated 
from Table 1 lie between 0.008 and 0.02 in approximate agreement with their 
position in Figs. 2 and 3.

In this section the luminosities determined above are used to compare with the 
luminosities other models predict in order to see if they agree with the core
masses given by Karakas. It would have been more consistent to compare with 
tracks of the original models of Karakas but she has not given post AGB tracks
for these models.

The effective temperature and the luminosity of the central stars shown in 
Table\,4 are plotted as points in Fig.\,4. The individual PN is identified on
the left side of the diagram. Several different models are plotted as solid 
lines on the figure. The core mass and initial heavy element abundance Z used 
in the model is shown in the diagram. Not shown on the diagram are the
phase of the helium shell flash cycle at which the ejection is assumed to 
occur and the mass loss rate used in making that model. The mass loss rate in
particular is a very poorly known quantity and can result in large 
uncertainties in the models. In Fig.\,4 the results of Vassiliadis \& Wood 
\cite{wood} are used for the core mass M=0.57M\smallsun, M=0.67M\smallsun~and 
0.91M\smallsun~models (the first has Z=0.016, the two other models have 
Z=0.008), and the results of Sch\"{o}nberner \cite{schon} for the 
M=0.55M\smallsun~ model. 

As can be seen in the diagram, two of the three PNe (NGC\,6537 and NGC\,6302)
for which it was concluded 
on the basis of their abundance that they originate from high mass stars, 
indeed are in the high mass region of the HR diagram. The third PN (He2-111)
might also be in this region but the temperature determination is too uncertain
to be sure of this. The three other bipolar PNe, NGC\,2440, Hb5 and Hu1-2 whose
abundance indicated a somewhat lower core mass, are in a
rather higher mass portion of the HR diagram. The precise value of the initial
mass agrees less well, since on the basis of the abundances we predicted in 
Sec.2 that these PNe originated from an initial mass of 4M\smallsun~to 
5M\smallsun~which following Karakas \cite{karakas} corresponds to a core mass 
of slightly higher than 0.8M\smallsun~while in the HR diagram they are in the 
position of the 0.67M\smallsun
track. In addition several other PNe lie in the same position. NGC\,6886 has
a high N/O ratio and on this basis was suspected to originate from a higher
mass star. The bipolar PN NGC\,6445, which because of its high He/H and low
N/O ratios was considered enigmatic, is also in this region. But so are 
NGC\,7027 and NGC\,6741 whose abundances would place them in the group of 
low initial mass objects. Also enigmatic are NGC\,5315 and M1-42 which on the
basis of both high He/H and N/O ratios might be expected to have somewhat 
higher core masses, are in the region of low mass objects on the HR diagram.

\begin{figure*}
 \centering
 \includegraphics[width=14cm,angle=90]{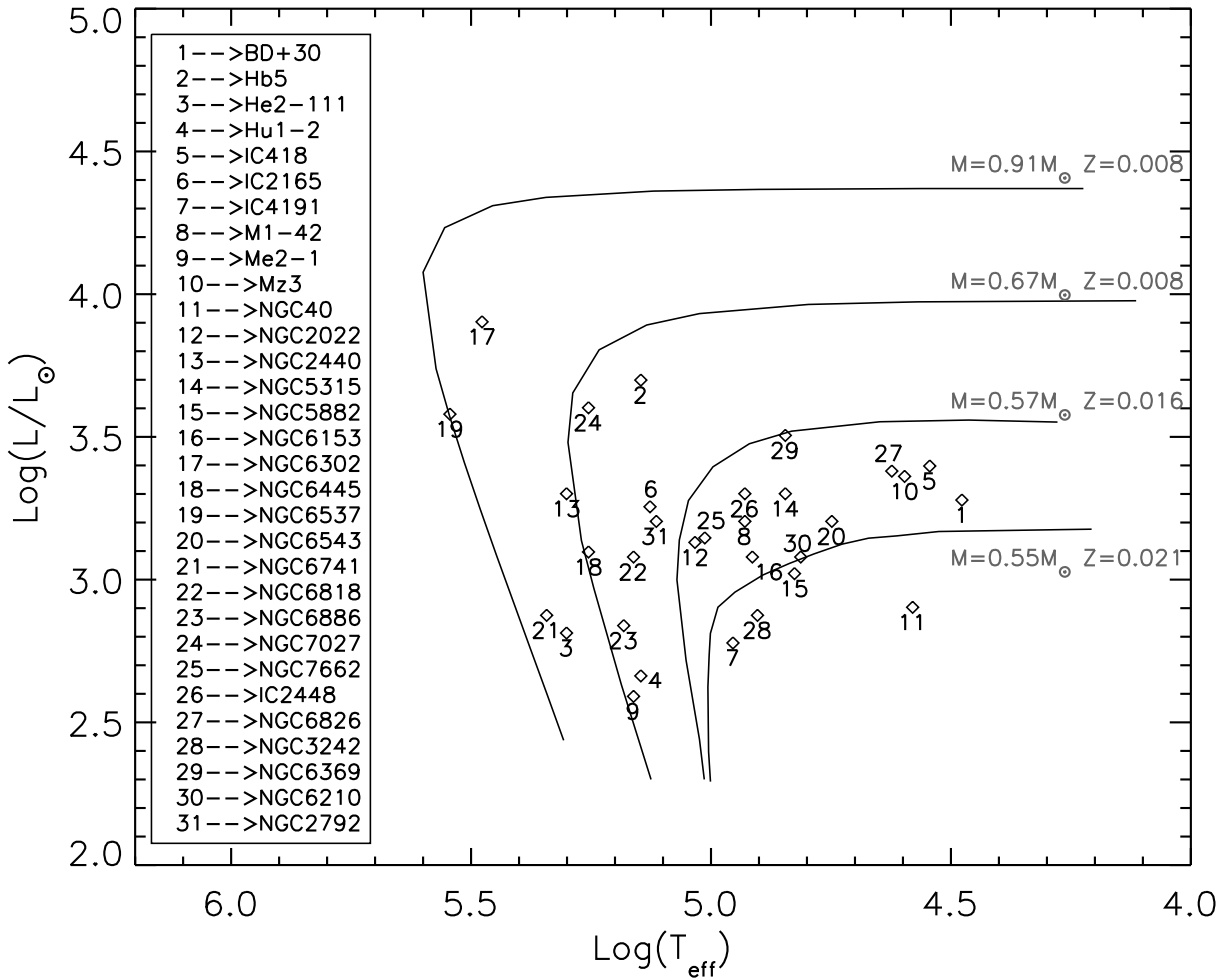}
 \caption{The HR diagram. The models for core masses of 0.57, 0.67 and 
0.91M\smallsun~are taken from Vassiliadis \& Wood\cite{wood} while the 
0.55M\smallsun~model is from Sch\"{o}nberner\cite{schon}.}

 \label{fig-4}
\end{figure*}

On the other hand, the majority of PNe which on the basis of their abundances 
are thought to have evolved from low mass stars, do indeed fall in the low
mass region of the HR diagram. We have checked to see whether those PNe which
are carbon rich (C/O$\geq$1) have a special position on the HR diagram, but we
find them to be at positions indistinguishable from other low initial mass
objects.

To summarize, there is a qualitative agreement between the two ways of
approaching stellar evolution: by either looking at the nebular
abundances or the position of the central star on the HR diagram. The
three bipolar PNe whose high N/O and He/H leads to the prediction
(Karakas\,\cite{karakas}) that they have core masses of
0.85M\smallsun~to 0.95M\smallsun~are in a position on the HR diagram
that is consistent with the theoretical 0.9M\smallsun~tracks of
Vassiliadis \& Wood\,\cite{wood} or the 0.91M\smallsun~tracks of
Bl\"{o}cker\,\cite{blocker} (not shown in Fig.4). The other three bipolar
PNe with somewhat lower abundances and where the predictions of
Karakas lead us to expect a slightly lower core mass, probably between
0.8M\smallsun~and 0.88M\smallsun, show less good agreement with the
model tracks. In Fig.4 these PNe agree with the 0.67M\smallsun~tracks
of Vassiliadis \& Wood\,\cite{wood} but they also agree with one of
the 0.6M\smallsun~tracks of Bl\"{o}cker\,\cite{blocker}.

Regarding the elliptical PNe: many of the central stars fall near the
tracks of stars of core mass between 0.55M\smallsun~and
0.57\smallsun~on the HR diagram. Stars of these masses are not
expected to show abundance changes which are at present observable, in
agreement with the observations. Karakas \cite{karakas} predicts a
core mass of 0.6M\smallsun~for these PNe, more in line with the value
of 0.58M\smallsun~expected from the average measured white dwarf
masses. The reason that the observed luminosity does not agree with
that predicted from the models is not clear. If the observed
luminosity were too low, the most likely cause would be a possible
underestimate of the observed PNe distance. To obtain agreement with
the model luminosity the average distance would have to be increased
by 70\%. We regard this as unlikely for an average value, although it may
occasionally occur for individual PNe.The other possibility is that the
relationship between core mass and luminosity found for the low core
mass models is not entirely correct.

\section{Summary and conclusions}

We have determined the masses of a selection of 33 well-studied
nebulae PNe by comparing the observed nebular abundances with that
predicted by the evolutionary models of Karakas\,\cite{karakas}, which
is the first systematic study of the evolution of lower mass stars.
A secondary purpose is to see whether the masses determined in this way
are consistent with the evolutionary tracks computed by
Sch\"{o}nberner\,\cite{schon}, Bl\"{o}cker\,\cite{blocker} and Vassiliadis \&
Wood\,\cite{wood}. The abundances used in this comparison are helium,
oxygen, nitrogen and carbon. The morphology is also considered in this
comparison but in a simplified form. The PNe are divided into two
categories: bipolar and elliptical (including round nebulae). The
result of the comparison of He vs N/O (section 2.2) indicate four PNe
have core masses greater than 0.9M\smallsun; three of these nebulae
are bipolar (NGC6302, NGC6537 and He2-111). There are also four
nebulae which appear to have a slightly lower initial mass, between
0.85M\smallsun~and 0.9M\smallsun~and have also undergone hot-bottom
burning.  Again three of these PNe are bipolar (NGC2440, Hb5 and
Hu1-2). Thus 6 of the 7 bipolar nebular in our selection have
abundances which indicate that they originated from initially high
mass stars. The only bipolar nebula which appears to be an exception
is NGC6445 which has a high He/H ratio but a low N/O ratio which is
difficult to understand on the basis of the models of Karakas.

NGC5315 appears to be an elliptical PN with a high mass. There are 
three other elliptical nebulae which on the basis of a rather high N/O ratio
seem to originate from stars of rather high mass. These are NGC6153, NGC6886 
and NGC2392. A unlikely explanation is that these PNe are really bipolar seen 
edge on making them look round. Two of these PNe do not have a high He/H ratio 
however, making this interpretation appear too simplified. All the other 
elliptical PNe have N/O and He/H ratios which do not substantially differ from 
solar and therefore have a core mass of less than approximately 0.7M\smallsun, 
using the evolutionary models of Karakas\,\cite{karakas}. We cannot determine
the mass more precisely for these nebulae, except to say that in the models of
Karakas those PNe with a high carbon abundance (C/O$\geq$1) have the higher 
mass.

An HR diagram for these nebulae was then constructed in order to see if the 
position of the PNe on this diagram can confirm the conclusions drawn from the
comparison of the abundances with the predictions of Karakas. The effective 
temperature of the central stars is determined, 
usually with an error that is less than 10\%. The determination of the 
luminosity can be made in two ways. First by computing it from the temperature 
and radius of the central star, and secondly by finding it from the nebular 
emission in both the hydrogen and helium lines. The second method has the 
additional assumption that the nebula is 'optically deep' to ionizing radiation
of hydrogen or helium. We find that the luminosity determined from the ionized
helium line is usually approximately equal to the luminosity found from the 
central star, indicating that the nebula absorbs all radiation which can doubly
ionize helium. For about 30\% of the nebulae the hydrogen ionizing radiation 
gives a lower luminosity indicating that in these cases the nebula is 'optically thin' to this radiation. In this way a method has been found to determine 
which nebulae are `optically deep' to the various radiation fields. Probably 
the largest uncertainty in the 
luminosity is the distance determination so that a subsection is devoted to a
discussion of the distances used. The resultant HR diagram is shown in Fig.4.
The high masses found from the abundances for the six bipolar PNe are 
consistent with their position on the HR diagram. The position of the seventh 
bipolar PN, NGC6445, is consistent with it being a high mass object, leaving 
its rather low N/O ratio as a problem. The position of NGC6886 is consistent
with a rather high mass. Most of the elliptical PNe have positions
consistent with low core masses. The biggest problems are: 1)  NGC7027, whose
HR position indicates a high mass while its abundances give a low  mass; 2) 
NGC5315 whose position indicates a low mass while its
abundance gives a high mass. M1-42 is also a problem but its distance is very 
poorly known and some of the observations are not very good. NGC2392 has a
different problem since the luminosity determined from the star is improbably 
high. The effective temperature that we have determined for this central star
does not apply to the star measured in the visual. The most likely solution is
that this central star is a binary and the secondary, which is unseen in the 
visual, is of a much higher temperature and is responsible for the high degree 
of ionization found. The possible presence of a hotter star has been suspected
earlier; Ciardullo et al.\cite{ciar} have looked for it but have found only a 
very faint nearby star. If 
this is the source of ionization it cannot be a main sequence star because if 
this was so it would be too distant to be associated with the nebula. But the
ionization source could be a star which is too close to the bright star to be
observed.

In general it appears that the initial masses as determined from the observed
abundances in conjunction with the models of Karakas\,\cite{karakas} are 
consistent with the initial masses predicted using the evolutionary models of
Sch\"{o}nberner\,\cite{schon}, Bl\"{o}cker\,\cite{blocker} and Vassiliadis \& 
Wood\,\cite{wood}. But being consistent is only a first step and both the 
models and the observed abundances should both be improved. 
Furthermore the various cases which appear to give inconsistent results must be
understood before we can speak of agreement.

\end{document}